\documentclass[article,preprint,amsmath,amssymb,amsfonts,superscriptaddress,aps]{revtex4-2}
\usepackage{times,graphicx,caption,epstopdf,dcolumn,bm,color,braket,multirow,xfrac,etoolbox}
\usepackage{hyperref} 
\usepackage[newcommands]{ragged2e}
\usepackage{amsmath}

\renewcommand{\thefigure}{\arabic{figure}}
\usepackage[font=small,justification=justified,format=plain]{caption}
\DeclareCaptionLabelSeparator{dot}{. }
\setcitestyle{square} 

\begin{document}

\title{Charge ordered phases in the hole-doped triangular Mott insulator \textit{4Hb}-TaS$_2$}

\author{Junho Bang}
\altaffiliation{\texorpdfstring{These authors contributed equally to this work}{}}
\affiliation{Department of Physics, Yonsei University, Seoul 03722, Republic of Korea}
\author{Byeongin Lee}
\altaffiliation{\texorpdfstring{These authors contributed equally to this work}{}}
\affiliation{Department of Physics, Yonsei University, Seoul 03722, Republic of Korea}
\author{Hyungryul Yang}
\affiliation{Department of Physics, Yonsei University, Seoul 03722, Republic of Korea}
\author{Sunghun Kim} 
\affiliation{Department of Physics, Ajou University, Suwon 16499, Republic of Korea}
\author{Dirk Wulferding}
\affiliation{Center for Correlated Electron Systems, Institute for Basic Science, Seoul 08826, Republic of Korea}
\author{Doohee Cho} \email{dooheecho@yonsei.ac.kr}
\affiliation{Department of Physics, Yonsei University, Seoul 03722, Republic of Korea}

\maketitle

\textbf{ \textit{4Hb}-TaS$_2$ has been proposed to possess unconventional superconductivity with broken time reveral symmetry due to distinctive layered structure, featuring a heterojunction between a 2D triangular Mott insulator and a charge density wave metal. However, since a frustrated spin state in the correlated insulating layer is susceptible to charge ordering with carrier doping, it is required to investigate the charge distribution driven by interlayer charge transfer to understand its superconductivity. Here, we use scanning tunneling microscopy and spectroscopy (STM/S) to investigate the charge ordered phases of 1\textit{T}-TaS$_2$ layers within \textit{4Hb}-TaS$_2$, explicitly focusing on the non-half-filled regime. Our STS results show an energy gap which exhibits an out-of-phase relation with the charge density. We ascribe the competition between on-site and nonlocal Coulomb repulsion as the driving force for the charge-ordered insulating phase of a doped triangular Mott insulator. In addition, we discuss the role of the insulating layer in the enhanced superconductivity of \textit{4Hb}-TaS$_2$.}

\section{INTRODUCTION}

Charge-ordered phases are primarily observed in strongly correlated electron systems. These phases arise due to strong on-site and nonlocal Coulomb interactions, which redistribute charges and lead to the formation of superstructures. Exploring these phenomena is motivated by their competitive relationship with superconductivity when carrier doping is introduced~\cite{imada1998metal,lee2006doping,kohsaka2007intrinsic,da2015charge}. Significant progress has been made in the field due to intense interest in cuprate high-temperature superconductors, which have complex phase diagrams featuring magnetic ordering, checkerboard charge ordering, pseudogap, and other quantum phenomena. However, a comprehensive understanding of the impact of Coulomb interactions and lattice geometries other than square lattice on charge ordering and superconductivity in frustrated correlated electron systems is still incomplete. These systems are hypothesized to support a gapless spin liquid phase~\cite{anderson1973resonating,law20171t,ruan2021evidence} and chiral superconductivity~\cite{kallin2016chiral,profeta2007triangular}.

\captionsetup{justification=Justified}
\begin{figure}[htb!]
\includegraphics[scale=1]{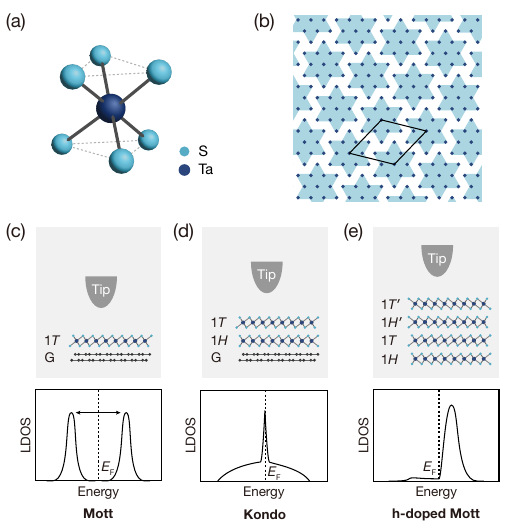}
\caption{Exotic quantum states arising from different layer stacking combinations of 1\textit{T}-TaS$_2$. (a) Schematics of the atomic structure of 1\textit{T}-TaS$_2$. A Ta atom (dark blue) is coordinated with six S atoms (bright blue) in an octahedral arrangement in the unit cell. (b)  $\sqrt{13}\times\sqrt{13}$ CDW pattern with Star of David reconstruction in 1\textit{T}-TaS$_2$. (c)-(e) Various quantum states and their corresponding spectral characteristics according to different stacking configurations of 1\textit{T} and 1\textit{H} layers. A Mott insulator (c), Kondo resonance (d), and hole-doped Mott insulator (e) observed on 1\textit{T}/Graphene, 1\textit{T}/1\textit{H}/Graphene, and 1\textit{T}/1\textit{H}/1\textit{T}/1\textit{H} vertical heterostructures, respectively.}
\label{Fig.1}
\end{figure}

 TaS$_2$, a transition metal dichalcogenides layered material, undergoes a charge density wave (CDW) transition at low temperatures. It is known to exhibit metallic behavior at higher temperatures due to the presence of a single valence electron in its 5$d$ orbital that stems from the covalent bonding between Ta and S atoms, conducting properties of CDW phases are differentiated according to the coordination of Ta and S atoms. In the monolayer, the 1\textit{H} polymorph remains as a metal, whereas the 1\textit{T} polymorph becomes a Mott insulator. 1\textit{T}-TaS$_2$, with octahedral coordination  (Fig.~\ref{Fig.1}(a)), forms Star of David (SD) clusters consisting of 13 Ta atoms, in which 12 non-bonding orbitals pair up and one non-bonding orbital resides at the center of the SD (Fig.~\ref{Fig.1}(b))~\cite{rossnagel2011origin}. Each SD cluster can be regarded as a unit cell where a pair of electrons can reside within. In the following we denote the number of electrons per non-bonding orbital at the center of the SD cluster as $n_{\rm SD}$ ($n_{\rm SD}=0$: unoccupied, $n_{\rm SD}=1$: half-filled, $n_{\rm SD}=2$: fully occupied). This configuration significantly enhances the on-site Coulomb repulsion, opening an energy gap near the Fermi level ($E_{\rm F}$)~\cite{fazekas1980charge}. With the Mott insulating character, a triangular lattice adds more dimension of complexity of the 1\textit{T}-TaS$_2$, accounting for spin frustration in non-bonding orbital sites~\cite{law20171t}.
 
\captionsetup{justification=Justified}  
\begin{figure*}[htb!]
\includegraphics[scale=1]{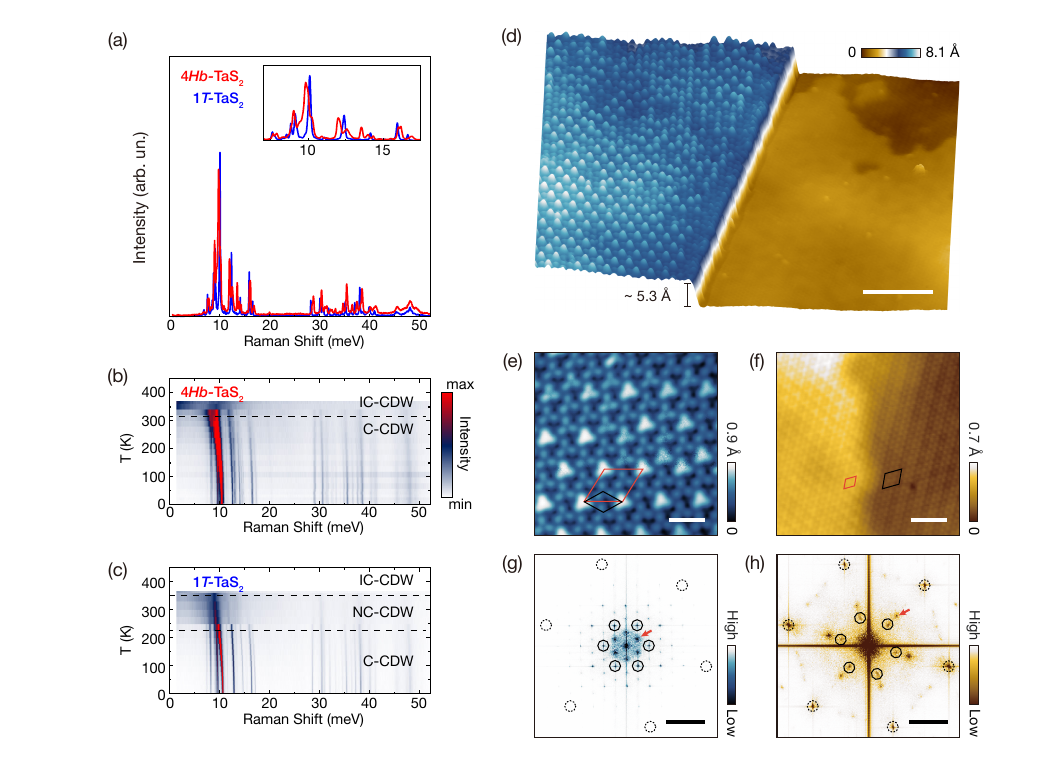}
\caption{Characterization and two distinct CDWs in 1\textit{T}- and 1\textit{H}-TaS$_2$. (a) Raman spectra of the \textit{4Hb}- and 1\textit{T} polymorphs of TaS$_2$ measured in the CDW phase at $T=4$ K. The inset zooms into the spectral range of CDW amplitude modes. (b), (c) Thermal evolution of Raman-active excitations in \textit{4Hb}- and 1\textit{T}-TaS$_2$, respectively, through various CDW transitions. (d) A 3D rendered STM image of \textit{4Hb}-TaS$_2$ with a monolayer step ($V_{\rm set}=-0.8$ V, $I_{\rm set}=50$ pA). The upper and lower terraces correspond to the typical surfaces of 1\textit{T} and 1\textit{H}-TaS$_2$ layers, respectively. Scale bar, $10$ nm. (e), (f) Zoomed-in STM images of the upper (e) and the lower (f) terrace. ($V_{\rm set}=-0.4$ V, $I_{\rm set}=100(50)$ pA). Scale bars for (e) and (f), $2$ nm. (g), (h) FFT of STM images of the upper (g) and lower (h) terrace. The dashed circles indicate the Bragg peaks of the pristine crystal lattice. The black circles correspond to the $\sqrt{13}\times\sqrt{13}$ and $3\times3$ CDW typically observed in bulk 1\textit{T} and 2\textit{H}-TaS$_2$, respectively. The red arrows highlight the peak of $\sqrt{3}\times\sqrt{3}$ of SD and $2\times2$ CDW that are not present in the bulk. Scale bars for (g) and (h), $1 \rm{nm^{-1}}$}\label{Fig.2}  
\end{figure*}

At its surface, bulk 1\textit{T}-TaS$_2$ can host two different insulating phases, a Mott and a band insulating phase, depending on distinct stacking orders of the surface terminating layers~\cite{lee2019origin,lee2021distinguishing,yang2023origin}. In the monolayer limit, it becomes a two-dimensional triangular Mott insulator~\cite{fazekas1980charge}, whose electronic structure can significantly depend on the substrate. Indeed, scanning tunneling spectroscopy (STS) measurements reveal distinct electronic structures of 1\textit{T}-TaS$_2$ surfaces deposited onto different substrates. A monolayer 1\textit{T}-TaS$_2$ grown on highly oriented pyrolytic graphite (HOPG) substrates shows electronic properties consistent with those of a correlated insulator (Fig.~\ref{Fig.1}(c)), which is characterized by two spectral peaks called upper and lower Hubbard bands~\cite{vano2021artificial}. Meanwhile, a 1\textit{T}/1\textit{H} bilayered heterostructure of TaS$_2$ on HOPG hosts an electron in the center orbital of the SD ($n_{\rm SD}=1$), thus localized spins in the 1\textit{T} layer are screened by itinerant electrons in the 1\textit{H} layer. This Kondo screening gives rise to the sharp zero-bias peak in the tunneling spectrum (Fig.~\ref{Fig.1}(d))~\cite{vano2021artificial,nayak2023first}. However on \textit{4Hb}-TaS$_2$, a 1\textit{T}/1\textit{H} heterostructure with the two polymorph layers stacked in an alternate sequence, the measurement exhibits a peak above $E_{\rm F}$ which corresponds to an unfilled narrow band centered at the SD ($n_{\rm SD}=0$). This feature can be attributed to either spectral weight transfer in a hole-doped Mott insulator~\cite{eskes1991anomalous,wang2020emergence} or pseudo-doping induced by weak hybridization between the 1\textit{T} and 1\textit{H} layer. (Fig.~\ref{Fig.1}(e))~\cite{wen2021roles}. These diverse electronic states emerging from different structure configurations indicate the interlayer coupling as a pivotal factor in the complexity of TaS$_2$-based heterostructures, and establish their potential for investigating novel quantum states of matter.

The difference in spectroscopic results between the Kondo-resonated 1\textit{T}/1\textit{H} bilayer and the hole-doped Mott insulator \textit{4Hb}-TaS$_2$ despite the structural similarity, implies that doping levels in the Mott-insulating 1\textit{T} layer can be modulated through the interlayer coupling. Controlling the carrier concentration leading to the transition from a Mott insulator to a superconducting phase is a widely utilized approach~\cite{lee2006doping}. It is also notable that intermediate fillings, between one and zero (or two) electron(s) per atomic site, give rise to metallic states characterized by unusual charge ordering and pseudogap features due to strong correlation~\cite{da2015charge}. With all conditions of the interlayer coupling and a frustrated structure, 1\textit{T} layer in \textit{4Hb}-TaS2 provides an ideal platform for investigating the effect of frustration on correlated phenomena.

In this study, we used scanning tunneling microscopy and spectroscopy (STM/S) at a low temperature to investigate the electronic structure and charge distribution of 1\textit{T} and 1\textit{H}-TaS$_2$ layers within the \textit{4Hb}-TaS$_2$ compound, particularly focusing on the triangular Mott insulator within the non-half filled regime. A new $\sqrt{3}\times\sqrt{3}$ SD superstructure superimposed on the conventional $\sqrt{13}\times\sqrt{13}$  superstructure is observed on the cleaved 1\textit{T} layer and the spatially resolved STS results show an energy gap with an out-of-phase relation with the local charge density. The observed charge-ordered insulating phase can be attributed to a bipolaronic insulating phase of the triangular Mott insulator in the intermediate hole-doping regime. Our results lead to a deeper understanding of the charge distribution and the emergent quantum phenomena in non-half-filled correlated electron systems with geometrically frustrated lattices.

\captionsetup{justification=Justified} 
\begin{figure*}[t]
\includegraphics[scale=1]{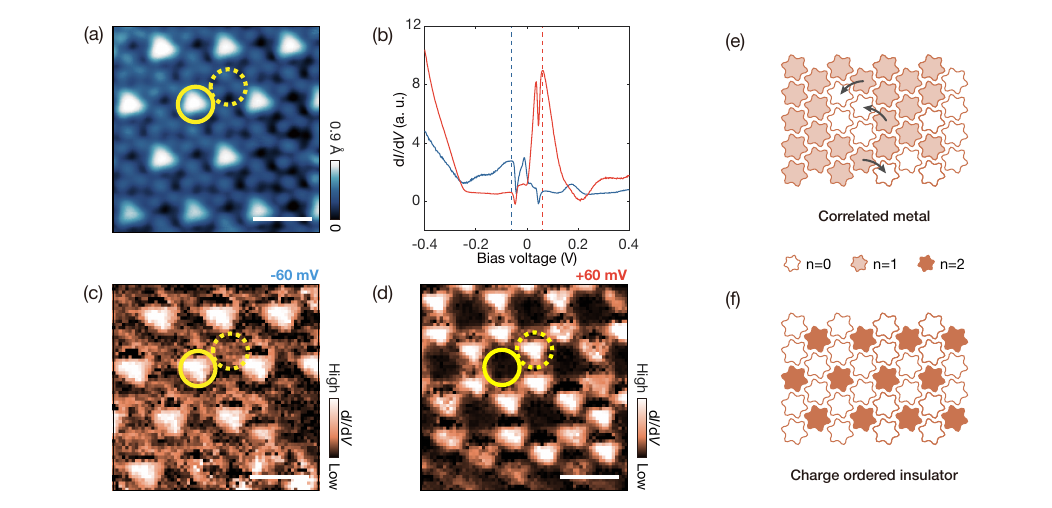}
\caption{Spectroscopic results and schematics of the charge ordered phase. (a) STM image of the $\sqrt{3}_{\rm~SD}\times\sqrt{3}_{\rm~SD}$ charge ordered phase at filled state ($V_{\rm set}=-0.4$ V and $I_{\rm set}=50$ pA). (b) Spatially averaged differential conductance $\mathrm{d}I/\mathrm{d}V$ spectra. Blue and red curves are acquired in brighter and darker sites in (a), respectively. ($V_{\rm set}=-0.4$ V, $I_{\rm set}=200$ pA, and $V_{\rm mod}=5$ mV). (c) and (d) $\mathrm{d}I/\mathrm{d}V$ maps at the energy of -60 mV (c) and +60 mV (d) which correspond to the dashed lines in (b) respectively. The solid and dashed yellow circles represent two types of CDW protrusions shown as brighter and darker sites in (a). Scale bars for (a), (c), and (d), $2$ nm. (e) A schematic of randomly distributed half-filled ($n_{\rm SD}=1$) and fully empty ($n_{\rm SD}=0$) SD. The black arrows indicate the potential hopping processes driven by adding voids in the half-filled insulator, otherwise strongly suppressed. (f) Charge redistribution of a hole-doped triangular Mott insulator with average filling factor $n_{\rm avg}=2/3$, the same as in the case shown in (e).}\label{Fig.3}  
\end{figure*}

\section{Experimental Details}
We used a commercial STM (UNISOKU-USM1200) for STM/S measurements. The experimental data were acquired at $T$ = 4.2 K in an ultra-high vacuum (UHV) environment with a pressure of $1\times10^{-10}$ Torr. Commercial TaS$_2$ crystals (HQ graphene) were grown using the chemical vapor transport (CVT) method with iodine as a transport agent. Bulk TaS$_2$ was cleaved at room temperature and transferred to the STM head, which was cooled to 4.2 K. STM tips were fabricated through electro-chemical etching of a tungsten wire and subsequently cleaned by electron beam heating under UHV conditions. The tip was subsequently characterized on a clean Au(111) surface~\cite{supplementalmaterial}. The STM images were obtained in constant current mode by applying a bias voltage ($V_{\rm set}$) to the sample. We employed a standard lock-in technique to obtain differential conductance (d$I$/d$V$) with AC voltage modulation ($V_{\rm mod}$ = 5 mV and $f_{\rm mod}$ = 613 Hz) added to the DC sample bias.
Comparative Raman scattering measurements on single crystalline samples of 1\textit{T}-TaS$_2$ and \textit{4Hb}-TaS$_2$ were carried out using a triple-stage spectrometer (Princeton Instruments TriVista) and a $\lambda = 561$ nm laser focused onto the sample with a spot diameter of 2 $\mu$m and a laser power of 0.1 mW. The samples were cooled via an open-flow He cryostat (Oxford MicroStat HR).

\section{RESULTS}
We confirm the \textit{4Hb}-phase via temperature-dependent Raman spectroscopy. This technique allows us to clearly distinguish \textit{4Hb}- from 1\textit{T}-polymorphs~\cite{nakashizu1984raman}, as both feature their own characteristic phonon spectrum (Fig.~\ref{Fig.2}(a)) as well as a distinct thermal evolution. Whereas 1\textit{T}-TaS$_2$ is characterized by a series of successive CDW phase transitions (commensurate to nearly-commensurate around  $T = 225$ K, nearly-commensurate to incommensurate around $T = 355$ K  upon heating), \textit{4Hb}-TaS$_2$ only has one phase transition at $T = 315$ K (Fig.~\ref{Fig.2}(b) and Fig.~\ref{Fig.2}(c)).

On the surface of cleaved \textit{4Hb}-TaS$_2$, two different polymorphs can be seen across a monolayer step, as shown in Fig.~\ref{Fig.2}(d). The step height is approximately 5.3 \AA, corresponding to the thickness of a TaS$_2$ monolayer~\cite{nayak2021evidence}. We identify the polymorphic structure of each surface by examining the CDW phases~\cite{ekvall1997atomic}. A zoomed-in STM topographic image of the upper terrace (Fig.~\ref{Fig.2}(e)) exhibits a triangular lattice made up of triangular-shaped protrusions. On the other hand, on the lower terrace, we observe the typical 3-fold stripe patterns of 1\textit{H}-TaS$_2$ (Fig.~\ref{Fig.2}(f)). The wavevectors of the typical CDWs of each polymorph are marked by black circles in the Fourier-transformed images of Fig.~\ref{Fig.2}(g) and~\ref{Fig.2}(h), well consistent with the $\sqrt{13}\times\sqrt{13}$ CDWs and the $3\times3$ CDWs of 1\textit{T}-TaS$_2$ and 1\textit{H}-TaS$_2$~\cite{wilson1975charge}, respectively.

In the 2D fast Fourier transform (FFT) of STM images of both the 1\textit{T} and 1\textit{H} layers, additional superstructures are present, highlighted by red arrows in Fig.~\ref{Fig.2}(g) and~\ref{Fig.2}(h), which are associated with the ground state of each polymorph. In particular, in the topographic image of the upper terrace (Fig.~\ref{Fig.2}(e)), two distinguished types of protrusions, one brighter and one darker, are regularly arranged in a $\sqrt{3}\times\sqrt{3}$ pattern on the SD lattice ($\sqrt{3}_{\rm SD}\times\sqrt{3}_{\rm SD}$). This $\sqrt{3}_{\rm SD}\times\sqrt{3}_{\rm SD}$ ordering is not fully commensurate due to the domain walls or random charge distribution, resulting in somewhat blurry peaks in Fig.~\ref{Fig.2} (g).Notably, the $\sqrt{3}_{\rm SD}\times\sqrt{3}_{\rm SD}$ order in 1\textit{T}-TaS$_2$ is absent when each SD is half-filled ($n_{\rm SD}=1$)~\cite{vano2021artificial} or fully unoccupied ($n_{\rm SD}=0$)~\cite{eskes1991anomalous}. The lower terrace shown in Fig.~\ref{Fig.2}(f) is divided into two distinct CDW regions of a $2\times2$ order with a brighter contrast and a $3\times3$ order with a darker contrast. A $3\times3$ order mainly appears on 2\textit{H}-TaS$_2$ as well as on the 1\textit{H} layer of \textit{4Hb}-TaS$_2$, while a $2\times2$ order has been reported for the electron-doped~\cite{hall2019environmental} or strained~\cite{gao2018atomic} 2\textit{H}-TaS$_2$. As shown in Fig. ~\ref{Fig.2}(d), our sample appears to be more corrugated than a typical sample~\cite{wen2021roles}. The strain might be critical in letting our sample host heterogeneous charge distribution~\cite{supplementalmaterial}.

Next, we turn to spatially resolved STS measurements on the 1\textit{T} surface to investigate the origin of the $\sqrt{3}_{\rm SD}\times\sqrt{3}_{\rm SD}$ charge ordering. Figure 3(a) illustrates a distinct pattern of alternating brighter and darker SDs, marked by solid and dashed yellow circles, respectively. These variations reflect the different electron fillings on the surface. The $\mathrm{d}I/\mathrm{d}V$ spectra presented in Fig.~\ref{Fig.3}(b) were obtained by spatially averaging each type of CDW protrusions. Contrary to the measured spectrum on the CDW protrusion in a monolayer 1\textit{T}-TaS$_2$ on graphene, which typically shows a gap with peaks around $-0.2$ and $+0.2$ eV, our measurements reveal that both brighter and darker sites on the 1\textit{T} surface exhibit distinct, single peaks in the filled and empty states, as indicated by the red and blue dashed lines in Fig.~\ref{Fig.3}(b). The red spectrum corresponds to that of a typical \textit{4Hb}-TaS$_{2}$~\cite{wen2021roles}, while the blue one is consistent with that of electron-doped 1\textit{T}-TaS$_{2}$~\cite{lee2021distinguishing,lee2020honeycomb}. The coexistence of these two distinct spectra has not been reported in \textit{4Hb}-TaS$_2$. This unique spectral characteristic suggests an energy gap opening, possibly linked to charge redistribution, significantly deviating from the electronic structure of monolayer 1\textit{T}-TaS$_2$. Meanwhile, there are dips at $\pm50$ mV in the spectra that exhibit slightly negative conductance. These could be due to resonant tunneling of localized states of tip and sample~\cite{lyo1989negative,supplementalmaterial}, or stem from electron tunneling coupled to the characteristic vibrational modes~\cite{weiss2010imaging,yin2020clarifying}. Note that they do not affect our interpretation of the relation between the filling factors and spectral features. 

\captionsetup{justification=Justified}
\begin{figure}[t!]
\includegraphics[scale=1]{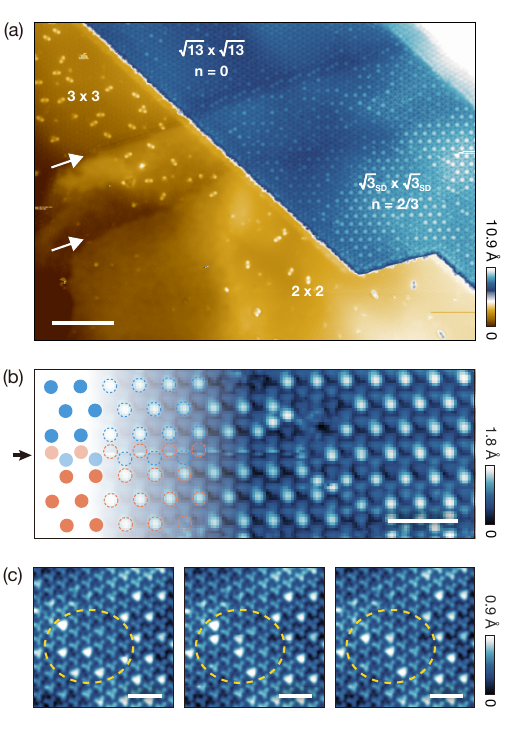}
\caption{Heterogeneous charge ordered phase and its dynamical properties. (a) An STM image of 1\textit{T} and 1\textit{H} surfaces containing multiple phases. The white arrows highlight the boundaries of distinct phase domains of the 1\textit{H} layer underneath the 1\textit{T} layer. Scale bar, $10$ nm (b) An STM image of $\sqrt{3}_{\rm SD}\times\sqrt{3}_{\rm SD}$ charge ordering with distinct CDW domains ($V_{\rm set}=-0.8$ V and $I_{\rm set}=50$ pA). Brighter sites are highlighted with circles and the color of the circles (blue and orange) represents the CDW phase of each domain. The black arrow points to the CDW domain wall between two domains and its fluctuations. Scale bar, $5$ nm. (c) A temporal-sequential STM image series of the same field of view ($V_{\rm set}=-2.0$ V and $I_{\rm set}=100$ pA). The yellow dashed circles highlight areas showing current-induced fluctuations. The time interval between the STM images is about 40 min. Further investigations using high-frequency measurements with a local probe will be required to reveal the origin of the dynamic behavior and its temporal characteristics. Scale bar, $3$ nm.}\label{Fig.4}  
\end{figure}

The $\mathrm{d}I/\mathrm{d}V$ maps acquired on the same region corresponding to Fig.~\ref{Fig.3}(a) unveil the origin of the charge redistribution. By comparing the $\mathrm{d}I/\mathrm{d}V$ maps at the energy levels of $-0.06$ eV and $+0.06$ eV, indicated by dashed lines in Fig.~\ref{Fig.3}(b), we observe a contrast inversion of the spatial $\sqrt{3}_{\rm SD}\times\sqrt{3}_{\rm SD}$ superstructure (Fig.~\ref{Fig.3}(c) and (d)). Furthermore, our two distinct spectra are consistent with the cases of fully empty ($n_{\rm SD}=0$) and fully filled ($n_{\rm SD}=2$) in the doped Mott-Hubbard systems, featured by spectral weight transfer into in-gap states with suppression of the Hubbard bands~\cite{eskes1991anomalous}. The electron doping forms the in-gap features below E$_{F}$, while the hole doping generates those above E$_{F}$. Our spectra show only a single peak, with no indications of any residual spectral features from the Hubbard band. Taken together, we may estimate the overall electron filling in the 1\textit{T}-TaS$_2$ layer to be $n_{\rm avg}=2/3$, where $n_{\rm avg}$ is the average number of electrons per SD cluster in the layer. This electron occupation differs from the half-filled ($n_{\rm avg}=1$) state of monolayer 1\textit{T}-TaS$_2$~\cite{vano2021artificial} and the fully empty ($n_{\rm avg}=0$) state of typical 1\textit{T} layers in \textit{4Hb}-TaS$_2$~\cite{eskes1991anomalous} that exhibit a commensurate $\sqrt{13}\times\sqrt{13}$ CDW. In our case, the average amount of charge transfer to the adjacent 1\textit{H} layer is not 1 but 1/3 electron for each SD cluster. Such fractional charge filling can be susceptible to charge redistribution, forming fully occupied and unoccupied dangling bonds, due to on-site and nonlocal Coulomb interaction.

Band filling is intimately related to the various charge-ordered phases emerging from 1\textit{T}-TaS$_2$. A half-filled monolayer of 1\textit{T}-TaS$_2$ typically behaves as a correlated insulator (Fig.~\ref{Fig.1}(c)). However, the alternating stacking of 1\textit{T} and 1\textit{H} layers leads to an interlayer charge transfer due to the work function difference between the two polymorphs as the valence electrons of 1\textit{T} layers flow to the 1\textit{H} layers in \textit{4Hb}-TaS$_2$~\cite{wang2018surface}. In line with this, a 1\textit{T}-TaS$_2$ layer in \textit{4Hb}-TaS$_2$ shows a hole-doped Mott phase in which the spectral peak is formed just above $E_{\rm F}$ with very weak (almost absent) spectral intensity below $E_{\rm F}$ (Fig.~\ref{Fig.1}(e)). The strongly asymmetric spectral feature suggests that one electron per $\sqrt{13}\times\sqrt{13}$  SD cluster is transferred to the 1\textit{H}-layer. Interestingly, even though the CDW is susceptible to band filling, the commensurate CDW in 1\textit{T}-TaS$_2$ is preserved in two different limits: half-filled ($n_{\rm avg}=1$) and fully empty ($n_{\rm avg}=0$).

In the intermediate doping regime, we can expect a charge rearrangement in a Mott insulator due to a strong correlation effect. For example, when a Mott insulator is charge depleted and thus its overall charge density approaches $n_{\rm avg}=2/3$, one can expect the half-filled ($n_{\rm SD}=1$) and the fully-empty ($n_{\rm SD}=0$) SD to coexist with a commensurate CDW within 1\textit{T} layers (Fig.~\ref{Fig.3}(e)). These voids allow for sequential electron hopping, which is strongly suppressed in the half-filled condition~\cite{eskes1991anomalous}. However, this correlated metallic state was not observed in our measurements. Instead, we suggest that electrons can be redistributed to form one SD cluster occupied by two electrons with its nearest neighbor clusters in the $\sqrt{3}_{\rm SD}\times\sqrt{3}_{\rm SD}$ unit cell remaining unoccupied (Fig.~\ref{Fig.3}(f)). Therefore, this demonstrates that a doped Mott insulator undergoes a CDW transition rather than becoming a correlated metal within a certain doping concentration range. 

It is evident from Fig.~\ref{Fig.4}(a) that the heterogeneous charge-ordered phase in the upper 1\textit{T}-TaS${\rm _2}$ layer is affected by the different CDW domains of the 1\textit{H}-TaS${\rm _2}$ below. Although we cannot explicitly determine whether the charge ordering observed in the exposed part of the 1\textit{H}-TaS${\rm _2}$ layer persists below the 1\textit{T}-TaS${\rm _2}$ layer, we can clearly observe that the domain walls separating the CDW phases in each layer overlap with each other. We can distinguish three domains. The first one is composed of fully empty SDs with $\sqrt{13}\times\sqrt{13}$ ordering that are aligned with the $3\times3$ CDW domain in the 1\textit{H}-layer below (upper region of Fig.~\ref{Fig.4}(a)), The second one exhibits a $\sqrt{3}_{\rm SD}\times\sqrt{3}_{\rm SD}$ reconstruction on top of a $2\times2$ CDW domain in the 1\textit{H}-layer (lower region of Fig.~\ref{Fig.4}(a)). On occasion, domain walls are present between $\sqrt{3}_{\rm SD}\times\sqrt{3}_{\rm SD}$ domains with fluctuations (Fig.~\ref{Fig.4}(b)). The third domain consists of randomly distributed fully filled SDs with a low concentration above a disordered $3\times3$ CDW domain in the 1\textit{H}-layer (middle region of Fig.~\ref{Fig.4}(a)). We can argue that a $2\times2$ CDW has a higher work function than the $3\times3$ CDW, and thus attracts fewer electrons from the 1\textit{T}-layer, assuming the absence of intercalation. These observations imply that interlayer charge transfer can be modulated by the heterogeneous CDW substrate, and the band-filling is crucial to determining the charge order in a triangular Mott insulator.

\section{DISCUSSION}

A similar charge-ordered phase has been observed in triangular arrays of group-IV adatoms on semiconductor surfaces at low temperatures~\cite{carpinelli1996direct,carpinelli1997surface}. Each metallic atom passivates the dangling bonds of the semiconductor surface and hosts one valence electron. This metallic electron configuration makes the systems susceptible to on-site and nonlocal Coulomb repulsion~\cite{profeta2007triangular}. In case of Pb on Si(111) and Ge(111) surfaces, charges are redistributed to form $\sqrt{3}\times\sqrt{3}$ superstructures at low temperatures~\cite{carpinelli1996direct,adler2019correlation}. The $n_{\rm Pb}= 0$, $1$, and $2$ states alternatingly reside in atomic sites within the half-filled ($n_{\rm avg}=1$) regime~\cite{cortes2013competing}. On the other hand, Sn on Si(111) does not exhibit charge ordered phase while it becomes superconducting upon increased doping concentration~\cite{ming2017realization,wu2020superconductivity}. The insulator-to-metal transition accompanied by a spectral weight transfer and a putative unconventional superconductivity~\cite{wolf2022triplet,biderang2022topological} are the aspects that remain ambiguous in the doped 1\textit{T}-TaS$_{2}$.  

Deviations from the half-filled case also lead to charge redistributions governed by competition between on-site and nonlocal Coulomb repulsion~\cite{watanabe2005charge,davoudi2008competition,tocchio2014phase}. Previous studies together with our observations on \textit{4Hb}-TaS$_2$ indicate that each SD strongly prefers only two discrete charged states, $n_{\rm SD}= 0$ or $2$, while the half-filled SDs ($n_{\rm SD}=1$) and current-induced fluctuations (Fig.~\ref{Fig.4}(c)) are only observed occasionally~\cite{nayak2023first}. The absence of a half-filled SD also indicates that our charge-ordered state is bipolaronic rather than paramagnetic. Since the CDW gap is larger than the Mott gap in 1\textit{T}-TaS$_2$, CDW is not altered even if an SD is fully empty or filled. Notably, all of the stable charge distributions result in insulating states rather than metallic states.

Due to the interlayer charge transfer, 1\textit{T}-TaS$_2$ likely loses the exotic electronic properties expected in a half-filled triangular Mott insulator, which have been considered as a prerequisite for chiral superconductivity with broken time-reversal symmetry~\cite{ribak2020chiral,gao2020origin,persky2022magnetic,fischer2023mechanism}. Instead, in \textit{4Hb}-TaS$_2$, a two dimensional metallic 1\textit{H}-TaS$_2$ layer is encapsulated by charge ordered insulating 1\textit{T}-TaS$_2$ layers. Note that \textit{4Hb}-TaS$_2$ ($T_{\rm c}\sim~2.7$ K)~\cite{ribak2020chiral} exhibits a higher critical temperature than 2\textit{H}-TaS$_2$ ($T_{\rm c}\sim~0.8$ K)~\cite{navarro2016enhanced}. The origin of this enhanced superconductivity has yet to be firmly established. Recalling the structural similarity of monolayer FeSe on insulators~\cite{ge2015superconductivity}, our observation can turn one's attention to the interface between the two-dimensional superconducting layer and the charge-ordered insulating layer~\cite{wang2018surface}, the dimension-dependent pairing strength~\cite{navarro2016enhanced}, and the suppressed CDW in 1\textit{H}-TaS$_2$~\cite{wang2018surface} in order to understand the exotic superconductivity of \textit{4Hb}-TaS$_2$. 

\section{CONCLUSION}
In conclusion, our STM measurements on \textit{4Hb}-TaS$_2$ reveal the emergence of charge-ordered phases in strongly correlated electron systems in the intermediate doping regime. Our findings indicate the presence of a $\sqrt{3}\times\sqrt{3}$ superstructure on the SD lattice in the 1\textit{T}-TaS$_2$ layer which is a triangular Mott insulator at half-filled condition. The spatially resolved STS measurement on the charge-ordered state reveals unique characteristics that point toward an out-of-phase charge distribution, providing evidence for a CDW insulating phase. This suggests the presence of an intermediate state within the 1\textit{T}-TaS$_2$ layer, between half-filled and fully empty states, which has a commensurate $\sqrt{13}\times\sqrt{13}$ CDW phase. This intermediate state is likely due to charge transfer modulated by the CDW phases of the sublayer. Our findings shed light on the charge-ordered insulating states in a geometrically frustrated Mott insulator. They also provide a route to control the electronic properties of layered materials by regulating interlayer couplings.

\section{ACKNOWLEDGEMENT}
The authors acknowledge T. Benschop, B. Jang, Y. W. Choi for valuable discussions. This work was supported by the National Research Foundation of Korea (Grant No. 2017R1A5A1014862 (J.B., B.L., H.Y., and D.C.), 2020R1C1C1007895 (J.B., B.L., H.Y., and D.C.), and RS-2023-00251265 (J.B., B.L., H.Y. and D.C.), 2021R1A6A1A10044950 (S.K.), RS-2023-00285390 (S.K.), and RS-2023-00210828 (S.K.)), the Yonsei University Research Fund of 2019-22-0209 (J.B., B.L., H.Y. and D.C.), an Industry-Academy joint
research program between Samsung Electronics and Yonsei University (J.B., B.L., H.Y. and D.C.). D.W. acknowledges support from the Institute for Basic Science (IBS) (Grant No. IBS-R009-Y3).

\newpage
\section{Supplemental Material}
\setcounter{figure}{0}
\makeatletter 
\renewcommand{\thefigure}{S\@arabic\c@figure}
\makeatother

\captionsetup{justification=Justified}
\begin{figure}[htb!]
\includegraphics[scale=1]{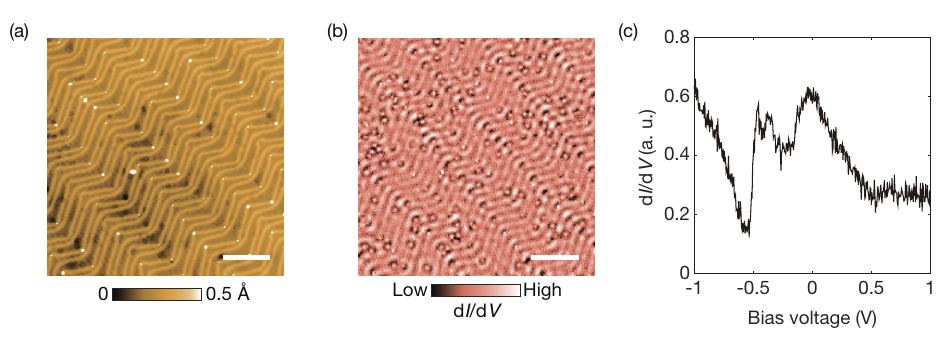}
\caption{STM tip characterizaion on Au(111) single crystal. (a) and (b) STM topographic image and simultaneously acquired $\mathrm{d}I/\mathrm{d}V$ map of clean Au(111) surface ($V_{\rm set}=-0.4$ V, $I_{\rm set}=100$ pA). Scale bars for (a) and (b), 20 nm. (c) $\mathrm{d}I/\mathrm{d}V$ spectrum of Au(111) surface. The peak around --0.5 eV corresponds to the surface state of Au(111) surface. The enhanced dip and peak features are possibly related to the localized states of an STM tip.}
\label{Fig.S1}
\end{figure}

\captionsetup{justification=Justified}
\begin{figure}[htb!]
\includegraphics[scale=1]{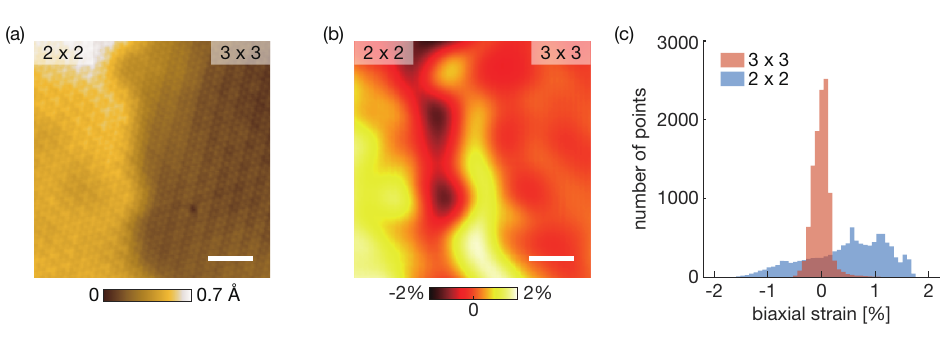}
\caption{Strain-induced heterogeneous CDW domains in the 1\textit{H} layer. (a) STM image of 1\textit{H}-layer showing $2\times2$ and $3\times3$ CDW domains. (b) Biaxial strain ($(s_{xx}+s_{yy})/2$) map calculated 
from the directional derivatives of the strain fields, $s_{xx}$ and $s_{yy}$. (c) Histogram of biaxial strain on the ($2\times2$) and the ($3\times3$) domains in the STM image in (a). Scales bars for (a) and (b), 2 nm.}
\label{Fig.S2}
\end{figure}


\begin{thebibliography}{50}

\bibitem{imada1998metal} M. Imada, A. Fujimori, and Y. Tokura, 
Metal-insulator transitions, 
Rev. Mod. Phys. \textbf{70}, 1039 (1998).

\bibitem{lee2006doping} P. A. Lee, N. Nagaosa, and X.-G. Wen, 
Doping a Mott insulator: Physics of high-temperature superconductivity, 
Rev. Mod. Phys. \textbf{78}, 17 (2006).

\bibitem{kohsaka2007intrinsic} Y. Kohsaka, C. Taylor, K. Fujita, A. Schmidt, C. Lupien, T. Hanaguri, M. Azuma, M. Takano, H. Eisaki, H. Takagi, \textit{et~al.}, 
An intrinsic bond-centered electronic glass with unidirectional domains in underdoped cuprates,
Science \textbf{315}, 1380 (2007).

\bibitem{da2015charge} E. H. da Silva Neto, R. Comin, F. He, R. Sutarto, Y. Jiang, R. L. Greene, G. A. Sawatzky, and A. Damascelli, 
Charge ordering in the electron-doped superconductor Nd$_{2-x}$Ce$_{x}$CuO$_{4}$, 
Science \textbf{347}, 282 (2015).

\bibitem{anderson1973resonating} P. W. Anderson, 
Resonating valence bonds: A new kind of insulator?, 
Mater. Res. Bull. \textbf{8}, 153 (1973).

\bibitem{law20171t} K. T. Law and P. A. Lee, 
1\textit{T}-TaS${_2}$ as a quantum spin liquid, 
Proc. Natl. Acad. Sci. U.S.A. \textbf{114}, 6996 (2017).

\bibitem{ruan2021evidence} W. Ruan, Y. Chen, S. Tang, J. Hwang, H.-Z. Tsai, R. L. Lee, M. Wu, H. Ryu, S. Kahn, F. Liou, \textit{et~al.}, 
Evidence for quantum spin liquid behaviour in single-layer 1\textit{T}-TaSe$_2$ from scanning tunnelling microscopy, 
Nat. Phys. \textbf{17}, 1154 (2021).

\bibitem{kallin2016chiral} C. Kallin and J. Berlinsky, 
Chiral superconductors, 
Rep. Prog. Phys. \textbf{79}, 054502 (2016).

\bibitem{profeta2007triangular} G. Profeta and E. Tosatti, 
Triangular Mott-Hubbard insulator phases of Sn/Si(111) and Sn/Ge(111) surfaces, 
Phys. Rev. Lett. \textbf{98}, 086401 (2007).

\bibitem{rossnagel2011origin} K. Rossnagel, 
On the origin of charge-density waves in select layered transition-metal dichalcogenides, 
J. Phys. Condens. Matter \textbf{23}, 213001 (2011).

\bibitem{fazekas1980charge} P. Fazekas and E. Tosatti, 
Charge carrier localization in pure and doped 1\textit{T}-TaS${_2}$, 
Physica B\&C \textbf{99}, 183 (1980).

\bibitem{lee2019origin} S.-H. Lee, J. S. Goh, and D. Cho, 
Origin of the insulating phase and first-order metal-insulator transition in 1\textit{T}-TaS${_2}$, 
Phys. Rev. Lett. \textbf{122}, 106404 (2019).

\bibitem{lee2021distinguishing} J. Lee, K. H. Jin, and H. W. Yeom, 
Distinguishing a Mott insulator from a trivial insulator with atomic adsorbates, 
Phys. Rev. Lett. \textbf{126}, 196405 (2021).

\bibitem{yang2023origin} H. Yang, B. Lee, J. Bang, S. Kim, D. Wulferding, S.-H. Lee, and D. Cho, 
Origin of Distinct Insulating Domains in the Layered Charge Density Wave Material 1\textit{T}-TaS${_2}$ Adv. Sci., 2401348 (2024).

\bibitem{vano2021artificial} V. Va{\v{n}}o, M. Amini, S. C. Ganguli, G. Chen, J. L. Lado, S. Kezilebieke, and P. Liljeroth, 
Artificial heavy fermions in a van der Waals heterostructure, 
Nature \textbf{599}, 582 (2021).

\bibitem{nayak2023first} A. K. Nayak, A. Steinbok, Y. Roet, J. Koo, I. Feldman, A. Almoalem, A. Kanigel, B. Yan, A. Rosch, N. Avraham, \textit{et~al.}, 
First Order Quantum Phase Transition in the Hybrid Metal-Mott Insulator Transition Metal Dichalcogenide 4\textit{Hb}-TaS$_2$, 
Proc. Natl. Acad. Sci. U.S.A. \textbf{120}, e2304274120 (2023).

\bibitem{eskes1991anomalous} H. Eskes, M. B. J. Meinders, and G. A. Sawatzky, 
Anomalous transfer of spectral weight in doped strongly correlated systems, 
Phys. Rev. Lett. \textbf{67}, 1035 (1991).

\bibitem{wang2020emergence} Y. Wang, Y. He, K. Wohlfeld, M. Hashimoto, E. W. Huang, D. Lu, S.-K. Mo, S. Komiya, C. Jia, B. Moritz, \textit{et al.},
Emergence of quasiparticles in a doped Mott insulator, 
Commun. Phys. \textbf{3}, 210 (2020).

\bibitem{wen2021roles} C. Wen, J. Gao, Y. Xie, Q. Zhang, P. Kong, J. Wang, Y. Jiang, X. Luo, J. Li, W. Lu, \textit{et~al.}, 
Roles of the Narrow Electronic Band near the Fermi Level in 1\textit{T}-TaS$_2$-Related Layered Materials, 
Phys. Rev. Lett. \textbf{126}, 256402 (2021).

\bibitem{supplementalmaterial} See Supplemental Material Section below for additional details on tip characterization and strain mapping.

\bibitem{nakashizu1984raman} T. Nakashizu, T. Sekine, K. Uchinokura, and E. Matsuura,
Raman study of charge-density-wave excitations in 4\textit{Hb}-TaS${_2}$, 
Phys. Rev. B \textbf{29}, 3090 (1984).

\bibitem{nayak2021evidence} A. K. Nayak, A. Steinbok, Y. Roet, J. Koo, G. Margalit, I. Feldman, A. Almoalem, A. Kanigel, G. A. Fiete, B. Yan, \textit{et~al.}, 
Evidence of topological boundary modes with topological nodal-point superconductivity, 
Nat. Phys. \textbf{17}, 1413 (2021).

\bibitem{ekvall1997atomic} I. Ekvall, J.-J. Kim, and \r{H}. Olin, 
Atomic and electronic structures of the two different layers in 4\textit{Hb}-TaS${_2}$ at 4.2 K, 
Phys. Rev. B \textbf{55}, 6758 (1997).

\bibitem{wilson1975charge} J. A. Wilson, F. Di Salvo, and S. Mahajan, 
Charge-density waves and superlattices in the metallic layered transition metal dichalcogenides, 
Adv. Phys. \textbf{24}, 117 (1975).

\bibitem{hall2019environmental} J. Hall, N. Ehlen, J. Berges, E. van Loon, C. van Efferen, C. Murray, M. R\"osner, J. Li, B. V. Senkovskiy, M. Hell, \textit{et~al.},
Environmental control of charge density wave order in monolayer 2\textit{H}-TaS$_2$, 
ACS Nano \textbf{13}, 10210 (2019).

\bibitem{gao2018atomic} S. Gao, F. Flicker, R. Sankar, H. Zhao, Z. Ren, B. Rachmilowitz, S. Balachandar, F. Chou, K. S. Burch, Z.Wang, \textit{et~al.}, 
Atomic-scale strain manipulation of a charge density wave, 
Proc. Natl. Acad. Sci. U.S.A. \textbf{115}, 6986 (2018).

\bibitem{lee2020honeycomb} J. Lee, K. -H. Jin, A. Catuneanu, A. Go, J. Jung, C. Won, S. -W. Cheong, J. Kim, F. Liu, H. -Y. Kee, \textit{et~al.}, 
Honeycomb-Lattice Mott Insulator on Tantalum Disulphide, 
Phys. Rev. Lett. \textbf{125}, 096403 (2020).

\bibitem{lyo1989negative} I.-W. Lyo, P. Avouris, 
Negative differential resistance on the atomic scale: implications for atomic scale devices, 
Science \textbf{245}, 1369-1371 (1989).

\bibitem{weiss2010imaging} C. Weiss, C. Wagner, C. Kleimann, M. Rohlfing, F.S. Tautz, and R. Temirov, 
Imaging Pauli Repulsion in Scanning Tunneling Microscopy, 
Phys. Rev. Lett. \textbf{105}, 086103 (2010).

\bibitem{yin2020clarifying} R. Yin, Y. Zheng, X. Ma, Q. Liao, C. Ma, B. Wang, 
Clarifying the intrinsic nature of the phonon-induced gaps of graphite in the spectra of scanning tunneling microscopy/spectroscopy, 
Phys. Rev. B \textbf{102}, 115410 (2020).

\bibitem{wang2018surface} Z. Wang, Y.-Y. Sun, I. Abdelwahab, L. Cao, W. Yu, H. Ju, J. Zhu, W. Fu, L. Chu, H. Xu, \textit{et~al.}, 
Surface-limited superconducting phase transition on 1\textit{T}-TaS${_2}$, 
ACS Nano \textbf{12}, 12619 (2018).

\bibitem{carpinelli1996direct} J. M. Carpinelli, H. H. Weitering, E. W. Plummer, and R. Stumpf, 
Direct observation of a surface charge density wave, 
Nature \textbf{381}, 398 (1996).

\bibitem{carpinelli1997surface} J. M. Carpinelli, H. H. Weitering, M. Bartkowiak, R. Stumpf, and E. W. Plummer, 
Surface charge ordering transition: $\alpha$ phase of Sn/Ge(111), 
Phys. Rev. Lett. \textbf{79}, 2859 (1997).

\bibitem{adler2019correlation} F. Adler, S. Rachel, M. Laubach, J. Maklar, A. Fleszar, J. Sch{\"a}fer, and R. Claessen, 
Correlation-driven charge order in a frustrated two-dimensional atom lattice, 
Phys. Rev. Lett. \textbf{123}, 086401 (2019).

\bibitem{cortes2013competing} R. Cort{\'e}s, A. Tejeda, J. Lobo-Checa, C. Didiot, B. Kierren, D. Malterre, J. Merino, F. Flores, E. G. Michel, and A. Mascaraque,
Competing charge ordering and Mott phases in a correlated Sn/Ge(111) two-dimensional triangular lattice, 
Phys. Rev. B \textbf{88}, 125113 (2013).

\bibitem{ming2017realization} F. Ming, S. Johnston, D. Mulugeta, T. S. Smith, P. Vilmercati, G. Lee, T. A. Maier, P. C. Snijders, and H. H. Weitering, 
Realization of a hole-doped Mott insulator on a triangular silicon lattice, 
Phys. Rev. Lett. \textbf{119}, 266802 (2017).

\bibitem{wu2020superconductivity} X. Wu, F. Ming, T. S. Smith, G. Liu, F. Ye, K. Wang, S. Johnston, and H. H. Weitering, 
Superconductivity in a hole-doped Mott-insulating triangular adatom layer on a silicon surface, 
Phys. Rev. Lett. \textbf{125}, 117001 (2020).

\bibitem{wolf2022triplet} S. Wolf, D. Di Sante, T. Schwemmer, R. Thomale, and S. Rachel, 
Triplet superconductivity from nonlocal Coulomb repulsion in an atomic Sn layer deposited onto a Si(111) substrate, 
Phys. Rev. Lett. \textbf{128}, 167002 (2022).

\bibitem{biderang2022topological} M. Biderang, M.-H. Zare, and J. Sirker, 
Topological superconductivity in \rm{Sn/Si}(111) driven by nonlocal Coulomb interactions, 
Phys. Rev. B \textbf{106}, 054514 (2022).

\bibitem{watanabe2005charge} H. Watanabe and M. Ogata, 
Charge Order and Superconductivity in Two-Dimensional Triangular Lattice at $n=2/3$, 
J. Phys. Soc. Japan. \textbf{74}, 2901 (2005).

\bibitem{davoudi2008competition} B. Davoudi, S. R. Hassan, and A. M. S. Tremblay, 
Competition between charge and spin order in the $t-U-V$ extended Hubbard model on the triangular lattice, 
Phys. Rev. B \textbf{77}, 214408 (2008).

\bibitem{tocchio2014phase} L. F. Tocchio, C. Gros, X.-F. Zhang, S. Eggert,
Phase diagram of the triangular extended Hubbard model, 
Phys. Rev. Lett. 113, 246405 (2014).

\bibitem{ribak2020chiral} A. Ribak, R. M. Skiff, M. Mograbi, P. Rout, M. Fischer, J. Ruhman, K. Chashka, Y. Dagan, and A. Kanigel, 
Chiral superconductivity in the alternate stacking compound 4\textit{Hb}-TaS$_2$, 
Sci. Adv. \textbf{6}, eaax9480 (2020).

\bibitem{gao2020origin} J. J. Gao, J. G. Si, X. Luo, J. Yan, Z. Z. Jiang, W. Wang, Y. Y. Han, P. Tong, W. H. Song, X. B. Zhu, Q. J. Li, W. J. Lu, Y. P. Sun, 
Origin of the large magnetoresistance in the candidate chiral superconductor 4\textit{Hb}-TaS$_2$, 
Phys. Rev. B \textbf{102}, 075138 (2020).

\bibitem{persky2022magnetic} E. Persky, A. V. Bj{\o}rlig, I. Feldman, A. Almoalem, E. Altman, E. Berg, I. Kimchi, J. Ruhman, A. Kanigel, and B. Kalisky,
Magnetic memory and spontaneous vortices in a van der Waals superconductor, 
Nature \textbf{607}, 692 (2022).

\bibitem{fischer2023mechanism} M. H. Fischer, P. A. Lee, and J. Ruhman, 
Mechanism for $\pi$ phase shifts in Little-Parks experiments: Application to 4\textit{Hb}- TaS$_2$ and to 2\textit{H}- TaS$_2$ intercalated with chiral molecules, 
Phys. Rev. B \textbf{108}, L180505 (2023).

\bibitem{navarro2016enhanced} E. Navarro-Moratalla, J. O. Island, S. Ma\~nas-Valero, E. Pinilla-Cienfuegos, A. Castellanos-Gomez, J. Quereda, G. Rubio-Bollinger, L. Chirolli, J. A. Silva-Guill{\'e}n, N. Agra{\"\i}t, \textit{et~al.}, 
Enhanced superconductivity in atomically thin TaS$_2$, 
Nat. Commun. \textbf{7}, 11043 (2016).

\bibitem{ge2015superconductivity} J.-F. Ge, Z.-L. Liu, C. Liu, C.-L. Gao, D. Qian, Q.-K. Xue, Y. Liu, and J.-F. Jia, 
Superconductivity above 100 K in single-layer FeSe films on doped SrTiO$_{3}$,
Nat. Mater. \textbf{14}, 285 (2015).

\end{thebibliography}
\end{document}